\documentclass[12pt]{iopart}
\pdfoutput=1
\expandafter\let\csname equation*\endcsname\relax
\expandafter\let\csname endequation*\endcsname\relax
\usepackage{amsmath}
\usepackage{amssymb}
\usepackage{bm,dsfont}
\usepackage{graphicx}
\usepackage{lastpage}
\usepackage{xcolor}
\usepackage{subfig}
\usepackage{dsfont} 
\usepackage[colorlinks=true,allcolors=black]{hyperref}  

\newcommand{\ket}[1]{\left |#1 \right \rangle}

\begin{document}
\graphicspath{{pictures/}}
\title[The effects of thermal and correlated noise on magnons]{The effects of thermal and correlated noise on magnons in a quantum ferromagnet}

\author{Jan Jeske$^{1,2}$, \'Angel Rivas$^{3,4}$, Muhammad H.~Ahmed$^1$, Miguel~A.~Martin-Delgado$^{3,4}$, Jared H.~Cole$^1$}

\address{$^1$ Chemical and Quantum Physics, School of Science, RMIT University, Melbourne 3001, Australia}

\address{$^2$ Fraunhofer Institute for Applied Solid State Physics, Tullastr. 72, 79108 Freiburg, Germany} 

\address{$^3$ Departamento de F\'isica Te\'orica, Facultad de Ciencias F\'isicas, Universidad Complutense, 28040 Madrid, Spain}

\address{$^4$ CCS-Center for Computational Simulation, Campus de Montegancedo UPM, 28660, Boadilla del Monte, Madrid, Spain}

\begin{abstract}
The dynamics and thermal equilibrium of spin waves (magnons) in a quantum ferromagnet as well as the macroscopic magnetisation are investigated. Thermal noise due to an interaction with lattice phonons and the effects of spatial correlations in the noise are considered. We first present a Markovian master equation approach with analytical solutions for any homogeneous spatial correlation function of the noise. We find that spatially correlated noise increases the decay rate of magnons with low wave vectors to their thermal equilibrium, which also leads to a faster decay of the ferromagnet's magnetisation to its steady-state value. For long correlation lengths and higher temperature we find that additionally there is a component of the magnetisation which decays very slowly, due to a reduced decay rate of fast magnons. This effect could be useful for fast and noise-protected quantum or classical information transfer and magnonics. We further compare ferromagnetic and antiferromagnetic behaviour in noisy environments and find qualitatively similar behaviour in Ohmic but fundamentally different behaviour in super-Ohmic environments.
\end{abstract}

\maketitle

\section{Introduction}
The investigation of spin waves and magnons has lead to the emerging field of magnonics \cite{Kruglyak2010, Lenk2011, Chumak2015} which aims to enable magnons as information carriers for both classical \cite{Khitun2010, Chumak2014} and quantum \cite{Khitun2001} information technology. Since magnons do not carry charge their interaction and dissipation is minimal compared to electronic circuits and thus magnonics could enable information processing with hugely reduced power consumption \cite{Chumak2015, Khitun2010}. Furthermore magnonics could enable transport \cite{Bose2003, Bose2007, Christandl2004, Ahmed2017}, processing \cite{Schneider2008, Vogt2014} and storage of quantum bits in the same platform of spin systems, since single-spin systems, such as the nitrogen- or silicon-vacancy centre in diamond, are strong candidates for future quantum bits. Magnons also enable new approaches to entanglement and quantum processing since (unlike photons) they can be created with a simple spin-gate operation on a spin. A rigorous investigation of their response to quantum noise including spatially correlated noise is therefore an important step towards further progress and miniaturisation of magnonic systems and devices.

Using quantum master equations to calculate macroscopic properties of solid state systems can yield details about the material properties and dynamics. This is a distinctly different application from the calculation of expectation values in small quantum systems of only a few states. For certain systems such as the quantum ferromagnet, analytical solutions can be obtained from master equation approaches. Numerical methods for large systems are possible via mapping of master equations to a quantum jump formalism \cite{Vogt2013}. The effects of uncorrelated noise in a quantum antiferromagnet have been investigated analytically \cite{Rivas2013} and shown the applicability and usefulness of this technique. In addition, master equations techniques have been applied in other condensed matter scenarios such as topologically ordered systems \cite{Viyuela2012, Rivas2013b, Rivas2017}. 

\begin{figure}[bth]
\centering
\includegraphics[width=\columnwidth]{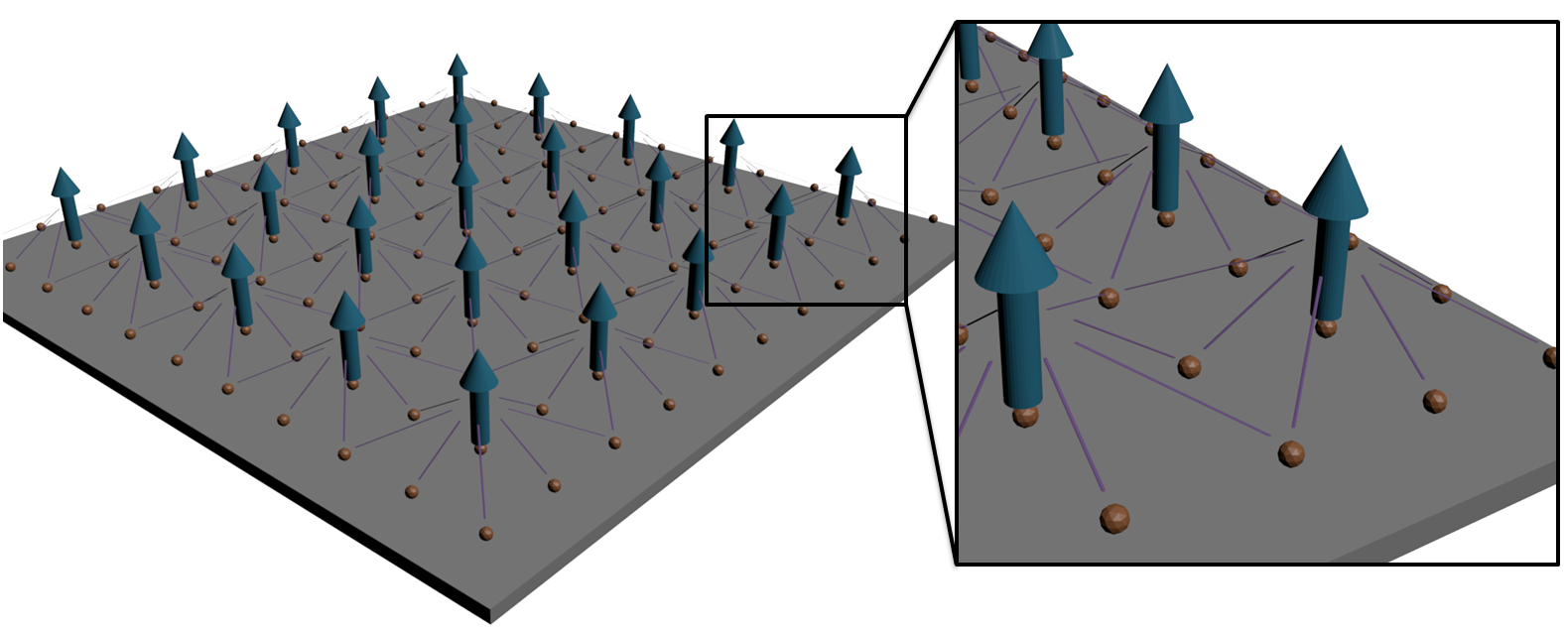}
\caption{Ferromagnetic spins (symbolised by arrows) interact with a noise environment of lattice phonons (symbolised by dots). Close spins couple to similar phononic sites and hence experience spatially correlated noise.}
\label{fig ferromagnet and phonons}
\end{figure}

The effects of spatially correlated noise are increasingly relevant for quantum technology devices, which aim for an increasing density of controlled quantum systems and thus the simplified model of uncorrelated noise are more likely to break down. Spatially correlated noise has shown to produce interesting and fundamentally different dynamics to uncorrelated noise, even in the Markovian regime. In ion traps for example the occurrence of decoherene-free subspaces has inspired quantum computation solutions \cite{Brown2003, Monz2009}.   
In the field of quantum metrology a re-instatement of the superior Heisenberg precision scaling has been proven possible in the presence of spatial noise correlations \cite{Jeske2014metrology}. In spin chains and light-harvesting complexes spatial noise correlations have shown to enable robust transport through protected states \cite{Jeske2013spinchain, Jeske2015, Ing2017}. 
These results pose the question how correlated noise affects the dynamics of magnons and macroscopic quantities in a quantum ferromagnet 

This paper is structured as follows: In section II we introduce the spin-wave Hamiltonian, in section III we discuss its interaction with a thermal environment, in section IV we derive and solve a master equation for magnons, in section V we discuss the macroscopic magnetisation, in section VI we point out relevant differences between ferro- and antiferromagnetic behaviour and reach our conclusions in section VII.

\section{Spin wave Hamiltonian}
We begin by introducing the key concepts and the Heisenberg model Hamiltonian for spins in real space with nearest-neighbour interaction. We then show how this Hamiltonian maps via the Holstein-Primakoff transform with a spin-wave approximation to a bosonic system. Subsequent transformation into $k$-space via Fourier lattice transform of the operators diagonalises the Hamiltonian and yields the dispersion relation of the system. This is the Hamiltonian describing `magnons', the elementary collective magnetic excitations. 

We start with a Heisenberg model for the spins in the quantum ferromagnet with only nearest-neighbour interaction and a uniform magnetic field $B$ in the negative $z$-direction: 
\begin{align}
H_{S}=J \sum_{\langle \textbf{r}, \textbf{r}'\rangle} \textbf{S}_\textbf{r} \cdot \textbf{S}_\textbf{r'} 
 -\sum_\textbf{r} \gamma B S^z_{\bf{r}} 
\label{Hamiltonian Heisenberg model for ferromagnet}
\end{align}
where $J<0$ for a ferromagnet. Antiferromagnetic behaviour ($J>0$) has been investigated similarly  \cite{Rivas2013}. Here $\gamma=g_e \mu_B/\hbar$ is the gyromagnetic ratio and $\textbf{S}_\textbf{r}=(S_x,S_y,S_z)$ is the vector spin-operator of at position $\textbf{r}=(x,y,z)$.

This Hamiltonian can be mapped to a system of interacting bosons via the Holstein-Primakoff transformation \cite{Holstein1940, Noltingmagnetism} for spins $S\geq 1/2$:
\begin{align}
\begin{aligned}
S^z_\textbf{r} &= S-n_\textbf{r}\\
S^+_\textbf{r} &= \sqrt{2S}\phi(n_\textbf{r}) a_\textbf{r}\\
S^-_\textbf{r} &= \sqrt{2S}a^\dagger_\textbf{r} \phi(n_\textbf{r})
\end{aligned}
\label{Holstein-Primakoff transform}
\end{align}
where $n_\textbf{r}=a^\dagger_\textbf{r} a_\textbf{r}$ and:
\begin{align}
\phi=\sqrt{1-\frac{n_\textbf{r}}{2S}}
\end{align}
This function is then approximated by a series expansion to first order in the normal-ordered number operator \cite{Noltingmagnetism}:
\begin{align}
\phi \approx 1-\left( 1-\sqrt{1-1/2S}\right) n_\textbf{r} \qquad \text{for }\;\langle a^\dagger_\textbf{r} a_\textbf{r} \rangle \ll 2S
\end{align}
We note that this linear spin-wave approximation limits the regime to low excitation numbers of magnons relative to their maximum set by the spin $2S$. This is fulfilled both in the ordered ferromagnetic equilibrium state as well as in the equilibration processes considered here, as long as the bath temperature is ``cold'' relative to the Debye temperature of the material, i.e.~such that the bath fluctuations cannot excite large numbers of magnons into the system. Furthermore, we will be working in a 3D system, leading to spontaneous symmetry breaking at finite temperature further justifying the spin-wave treatment to be qualitatively correct.

With the linear spin-wave approximation the Hamiltonian \eqref{Hamiltonian Heisenberg model for ferromagnet} becomes (for a $D$-dimensional system):
\begin{align}
H_{LSW}=E_0 + (\gamma B - 2J D S) \sum_{\textbf{r}} a^\dagger_\textbf{r} a_\textbf{r} 
+ J S \sum_{\langle \textbf{r}, \textbf{r'}\rangle} \left(a_\textbf{r}^\dagger a_{\textbf{r}'}+ a^\dagger_{\textbf{r}'} a_\textbf{r}\right) \label{eq H_lsw}
\end{align}
where the constant energy shift $E_0=J N D S^2 + \gamma B S N$ with the total number of spins $N$ originates from a summation over a constant. 

The Hamiltonian is then diagonalized for $D=3$ by replacing $a_\textbf{r}$ and $a^\dagger_\textbf{r}$  with their three-dimensional Fourier lattice transform \cite{Longophd, Noltingmanybody, Longo2011}:
\begin{align}
\begin{aligned}
a_\textbf{r}& =a_{x,y,z} = \frac{1}{\sqrt{N}} \sum_{\textbf{k}} e^{i \textbf{k}\cdot \textbf{r}} a_\textbf{k} \\
a_\textbf{r}^\dagger &= \frac{1}{\sqrt{N}} \sum_\textbf{k} e^{-i \textbf{k}\cdot \textbf{r}} a_\textbf{k}^\dagger 
\end{aligned} 
\label{eq. Fourier lattice transform}
\end{align}
Taking into consideration a simple cubic lattice as the discrete spin lattice, the detailed expression for the wave vector elements is $k_x=\frac{2\pi}{N_x d_x} n_x$ and $r_x=d_x \tilde n_x$, where $N_x$ is the number of sites in $x$-direction, $d_x$ is the lattice constant in $x$-direction and $n_x, \tilde n_x \in \mathds{N}$ are the integer summation indices. Analogous expressions apply to $y$- and $z$-directions. 
For the Hamiltonian, Eq.~\eqref{eq H_lsw}, we choose the three-dimensional nearest-neighbour parameterization $\sum_{\langle \textbf{r}, \textbf{r'}\rangle} \rightarrow \sum_{\bf{r}} \sum_{\mu=1}^3$ and $\textbf{r}'\rightarrow \textbf{r}+d_\mu \mathbf{\hat r_\mu}$, where $\bf{\hat r_\mu}$ is the unit vector in $x$-,$y$- or $z$-direction and $d_\mu$ the respective lattice constant. One then diagonalises the Hamiltonian in $k$-space using $  \sum_{\bf{r}} \exp\left[i (\textbf{k}-\textbf{k}')\cdot \textbf{r}\right] = \delta_{k_x,k'_x} \, \delta_{k_y,k'_y} \, \delta_{k_z,k'_z} $ and finds:
\begin{align}
H_{LSW}&=E_0+\sum_\textbf{k}\omega(\textbf{k}) a_\textbf{k}^\dagger a_\textbf{k}  \label{Hamiltonian ferromagnet diagonal in $k$-space}
\\
&\quad \text{with  } \quad\omega(\mathbf{k})=\left[\gamma B + 2 J S \sum_{\mu=1}^3 (\cos \,k_\mu d_\mu- 1) \right] \label{eq dispersion relation} 
\end{align}
This is the diagonal linear spin wave Hamiltonian. The excitations of these uncoupled harmonic oscillators are called magnons as they represent the elementary magnetic excitation. The $\mathbf{k}$-dependent energy of the magnons $\omega(\mathbf{k})$ defines their dispersion relation.

\section{Interaction with a thermal environment}
\label{sec Interaction with a thermal environment}
In this section we extend our model of the system of a ferromagnet to be an open quantum system, which allows us to model the interaction between the spins and the lattice phonons, which we will assume to be in a thermal equilibrium. Instead of assuming an ad-hoc master equation we will show a more rigorous treatment and start by assuming an interaction Hamiltonian between the ferromagnetic system and an environment of bosonic phonon-modes, Fig.~\ref{fig ferromagnet and phonons}. This is the basis for the derivation of a master equation. Any general interaction Hamiltonian \cite{Rivasbook} can always be written in the mathematical form $H_{int}=\sum_j s_j B_j$, where multiple system operators $s_j$ couple to environmental bath operators $B_j$. Defining these operators $s_j$ and $B_j$ in this form of the coupling is the starting point for a treatment with the Bloch-Redfield formalism, which allows us to write down the master equation directly from this form \cite{Jeske2013formalism, Jeskethesis, Bloch1957, Redfield1957}. 

We will start with an interaction Hamiltonian with spin-operators in real space and include  an arbitrary spatial correlation function $f(|\textbf{r}-\textbf{r}'|)$ of the system-environment couplings, which defines how the coupling between a spin and an environmental mode behaves as a function of distance $|\textbf{r}-\textbf{r}'|$ between them. We then show how this form simplifies by transformation into $k$-space such that magnons only couple to environmental modes with the same wave vector $\textbf{k}$. The spatial correlation function is then contained in the $\mathbf{k}$-dependent coupling constant. This simplification is essential in order to solve the resulting Bloch-Redfield equations analytically later. 

We note that introducing this spatial decay function $f(|\textbf{r}-\textbf{r}'|)$ in the system-environment \emph{couplings} is subtly different to the definition of a spatial decay function in the environmental \emph{noise correlations} as done elsewhere \cite{Jeske2013formalism}. In the latter case the decay function is bound by multipartite correlation rules while here the decay function of couplings is unrestricted in its functional form and will always result in a Lindblad form and a physical time evolution.

We assume that our system of spins couples to a large number of bosonic environmental modes, such as phonons. This coupling of the spins to the lattice phonons is the common spin-boson model \cite{Weissbook, Leggett1987, Unruh1995, DiVincenzo1995, Reina2002} where the spin operator $\mathbf{S}=(S^x, S^y, S^z)$ couples linearly to the lattice displacement operator $\mathbf{R}=(X,Y,Z)$. Details of an underlying microscopic model have been discussed \cite{Fransson2017} as a local exchange interaction between the electron spin and magnetic moment and the local couplings between the electronic charge and lattice displacements.  The interaction of the form $S^x X = S^x (A^\dagger + A)$, with the local position operator $X$ of the environmental mode, is one of the common spin-boson models \cite{Weissbook, Leggett1987}.  We chose this as the relevant example since an interaction $S^y Y$ yields completely analogous dynamics and an interaction $S^z Z$ would lead to terms which are negligible in the spin wave regime in analogy to the antiferromagnetic case \cite{Rivas2013}, see \ref{app:full spin-boson coupling} for details. We adopt capital letters for the environmental creation operators $A^\dagger_{\mathbf{r},j}$ of mode $j$ at spatial position $\mathbf{r}$ to distinguish them from the system modes $a^\dagger_\mathbf{r}$. At each position there is a collection of phonon modes with different energies $\omega_j$ and respective coupling strength $g(\omega_j)$: \\
\begin{align}
H_{int}=\sum_j g(\omega_j) \sum_{\textbf{r},\bf{r'}} f(|\textbf{r}-\textbf{r}'|)\, S^x_r (A^\dagger_{\textbf{r}',j} + A_{\textbf{r}',j})
\end{align}

After Holstein-Primakoff transformation with linear spin-wave approximation this interaction reads:
\begin{align}
H_{int}=\sqrt{2S}\sum_j g(\omega_j) \sum_{\textbf{r},\bf{r'}} f(|\textbf{r}-\textbf{r}'|)\, (a_{\bf{r}}^\dagger + a_\mathbf{r}) (A^\dagger_{\textbf{r}',j} + A_{\textbf{r}',j})
\end{align}
Using the secular approximation this can be simplified further. The secular approximation allows us to neglect coupling or noise terms (i.e.~off-diagonal superoperator elements) which are small relative to the difference of two on-diagonal superoperator elements). For a weak system-environment coupling $g(\omega_j)$ relative to the magnon energy we can neglect those terms that create both a magnon and an environmental phonon, i.e. terms proportional to $a^\dagger A^\dagger$ and $a A$, since these terms create small off-diagonal superoperator elements proportional to $g(\omega_j)$ corresponding to diagonal elements separated by the magnon energy. We only take into account terms where a magnon is created and an environmental phonon annihilated or vice versa, i.e.~$a^\dagger A$ and $a A^\dagger$ terms. This is more intuitively understood in the interaction picture of the system and environment where the $a^\dagger A^\dagger$ and $a A$ terms become fast oscillating terms with a frequency corresponding to the sum of magnon and phonon energy. The fast oscillation averages out their effect. The $a^\dagger A$ and $a A^\dagger$ terms are slow or non-oscillating terms since two counter-rotating factors cancel each other out. In this way the secular approximation can also be regarded as a rotating-wave approximation. Neglecting the $a^\dagger A^\dagger$ and $a A$ terms is a fairly common procedure in master equation approaches \cite{Breuerbook, Gardinerbook}, mathematically due to the Riemann-Lebesgue lemma \cite{Rivasbook}.
The interaction Hamiltonian becomes:
\begin{align}
H_{int}=\sqrt{2S}\sum_j g(\omega_j) \sum_{\textbf{r},\bf{r'}} f(|\textbf{r}-\textbf{r}'|)\, a_{\bf{r}}^\dagger A_{\textbf{r}',j} + \text{ h.c.} \label{interaction Hamiltonian real space}
\end{align}
We note that for this approximation to be generally correct the condition $g(\omega_j) \ll \gamma B$ is required since there are otherwise system magnons with an energy that is not large enough to justify the separate scales required for the secular approximation. In this case the exact form of the coupling becomes more relevant and $a^\dagger A^\dagger$ terms as well as $a A$ terms need to be considered. However, this condition can be fulfilled even for magnetic fields small compared to the spin-spin coupling $J \gg \gamma B \gg g(\omega_j)$. 

Again we make use of the Fourier lattice transform, eq.~\ref{eq. Fourier lattice transform}, and rewriting $e^{i(\mathbf{k'\cdot r'}-\mathbf{k \cdot r})}=e^{i\textbf{k}'\cdot(\mathbf{r'-r})} e^{i(\mathbf{k'-k})\cdot\textbf{r}}$ we find:
\begin{align}
H_{int}=\sqrt{2S} \sum_j g(\omega_j) \sum_{\textbf{k},\bf{k'}} \sum_{\textbf{r},\bf{r'}} e^{i\textbf{k}'\cdot(\mathbf{r'-r})} f(|\textbf{r}-\textbf{r}'|)\, 
\frac{1}{N}e^{i(\mathbf{k'-k})\cdot\textbf{r}} a_{\bf{k}}^\dagger A_{\textbf{k}',j} + \text{h.c.} 
\label{eq for appendix}
\end{align}
We then substitute $\bf{u}=\mathbf{r'-r}$ and write:
\begin{align}
\sum_{\mathbf{r'}=1}^N e^{i\textbf{k}'\cdot(\mathbf{r'-r})} f(|\mathbf{r-r}'|) 
=\sum_{\textbf{u}=-N/2}^{N/2-1} e^{i\textbf{k}'\cdot\mathbf{u}} f(|\textbf{u}|) 
\end{align}
or, to be more precise, we substitute for each component $r'_\mu-r_\mu=d_\mu \tilde n'_\mu-d_\mu \tilde n_\mu = d_\mu \hat n_\mu = u_\mu$, where $d_\mu$ is the lattice constant of dimension $\mu$, the variables $\tilde n_\mu, \tilde n'_\mu, \hat n_\mu \in \mathds{Z}$ are the integer summation indices and $\mu$ takes the values of the three dimensions:  
\begin{align}
\sum_{\tilde n'_\mu=1}^{N_\mu} e^{ik_\mu' d_\mu(n'_\mu-n_\mu)} f(|\mathbf{r-r}'|) 
= \sum_{\hat n_\mu=-N_\mu/2}^{N_\mu/2-1} e^{ik'_\mu d_\mu \hat n_\mu} f(|\textbf{u}|) 
\end{align}
In doing so we have assumed that the expression is independent of $\textbf{r}$, and that the summation over $\bf{r}'$ always runs over all possible values of the difference $\mathbf{r'-r}$. Physically this assumption means that all edge effects are neglected, which occur, when any $r_\mu$ is close to 1 or $N_\mu$. We then identify the remaining summation over r as a Kronecker-Delta and after performing the summation over $\textbf{k}'$, the interaction Hamiltonian then becomes (\ref{app: identifying the delta function}):
\begin{align}
H_{int}&=\sqrt{2S} \sum_j g(\omega_j) \sum_{\textbf{k}} 
\underbrace{ \sum_{\bf{u}} e^{i\textbf{k}\cdot\mathbf{u}} f(|\textbf{u}|)}_{F(\textbf{k})} 
\,a_{\bf{k}}^\dagger A_{\textbf{k},j}   + \text{h.c.} 
\label{eq for appendix 2}
\end{align}
This is the interaction Hamiltonian in $k$-space in which we find that magnons of the wave-vector $\bf{k}$ only couple to environmental modes of the same wave vector. The respective coupling strength is given by the three-dimensional Fourier transform of the coupling correlation function $f(|\mathbf{r-r}'|)$. This Fourier transform $F(\bf{k})$ is explicitly:
\begin{align}
F(\textbf{k})= \sum_{\hat n_x=N_x/2}^{N_x/2-1} \sum_{\hat n_y=N_y/2}^{N_y/2-1} \sum_{\hat n_z=N_z/2}^{N_z/2-1} f(|u|) 
e^{i(k_x d_x \hat n_x+k_y d_y \hat n_y +k_z d_z \hat n_z)} 
\label{correlation funct Fourier transform discrete}
\end{align}
For large $N$ and smooth $f(|u|)$ one can take the continuous limit and integrate. Since we neglected edge effects and assumed the difference $\bf{u}=\bf{r}-\bf{r}'$ to go over all possible values, this is the thermodynamic limit and integration should go from $-\infty$ to $\infty$, where the normalisation of $f(|u|)$ does not need to be adapted if it decays on a small scale relative to the crystal lengths $d_\mu N_\mu$.
\begin{align}
F(\textbf{k})&=\iiint d\hat n_x\, d\hat n_y \, d\hat n_z f(|u|) e^{i(k_x d_x \hat n_x+k_y d_y \hat n_y +k_z d_z \hat n_z)} \\
&=\frac{1}{d_x d_y d_z} \iiint du_x \, du_y\, du_z f(|u|)e^{ i \textbf{k}\cdot \textbf{u}}
\label{correlation function - Fourier transform F(k)}
\end{align}
where $V_d=d_x d_y d_z$ is the volume of the unit cell and $\textbf{u}$ carries the unit  length while $\hat n_\mu$ is dimensionless. 

Therefore, the Fourier transform of $f(|u|)$ depends only on the magnitude of the wave vector, $F(\mathbf{k})=F(k)$, allowing further simplifications, \ref{app simplifying Fourer transform F(k) to 1D integral}. We note that this isotropy follows from the assumed isotropy of the noise correlation function $f(|u|)$ for the whole lattice. It is not connected to any isotropy of the individual site's system-bath coupling operators.

With this the Hamiltonian is:
\begin{align}
H_{int}&=\sqrt{2S} \sum_j g(\omega_j) \sum_{\textbf{k}} F(k) 
\,a_{\bf{k}}^\dagger A_{\textbf{k},j}   + \text{h.c.}  
\label{eq H_int final}
\end{align}
This interaction Hamiltonian in $k$-space is much simpler as it only couples magnons of wave vector $\mathbf{k}$ to phonons of the same wave vector and thus there is only one summation over $\mathbf{k}$, despite two summations ocurring in the spatially correlated real-space form, Eq.~\eqref{interaction Hamiltonian real space}. We therefore use this as the starting point for our master equation. 

\section{Master equation and thermalisation of magnons}
For the standard Bloch-Redfield approach we can now identify the full Hamiltonian $H=H_{sys} + H_{int} + H_{env}$, where the system Hamiltonian $H_{sys}$ is given by the linear spin-wave Hamiltonian from eq.~\ref{Hamiltonian ferromagnet diagonal in $k$-space}, the interaction Hamiltonian $H_{int}$ is given by eq.~\ref{eq H_int final} and the environmental Hamitonian is defined as the bosonic environmental modes:  

\begin{align}
H=H_{sys} + H_{int} + &H_{env}\\
H_{sys}= & H_{LSW}=E_0+\sum_\textbf{k}\omega(\textbf{k}) a_\textbf{k}^\dagger a_\textbf{k}  
\\
&\qquad \text{with  } \quad\omega(\mathbf{k})=\left[\gamma B + 2 J S \sum_{\mu=1}^3 (\cos \,k_\mu d_\mu- 1) \right]  \\
H_{int}=&\sqrt{2S} \sum_j g(\omega_j) \sum_{\textbf{k}} F(k) 
\,a_{\bf{k}}^\dagger A_{\textbf{k},j}   + \text{h.c.} \\
H_{env}=&\sum_{\textbf{k},j} \omega_{\mathbf{k},j} \,A^{\dagger}_{\mathbf{k},j}A_{\mathbf{k},j}
\end{align}

The interaction Hamiltonian has the form $H_{int}=\sum_j s_j B_j$, mentioned above, i.e.~multiple products of system operators $s_j$ with bath operators $B_j$. We label the system operators $s_{1\mathbf{k}}=a_\textbf{k}^\dagger$ and Hermitean conjugate $s_{2\mathbf{k}}=a_\textbf{k}$ with the corresponding bath operators $B_{1\mathbf{k}}=\sqrt{2S}\sum_j g(\omega_j) F(k) A_{\textbf{k},j}$ and $B_{2\mathbf{k}}=\sqrt{2S}\sum_j g^*(\omega_j) F^*(k) A_{\textbf{k},j}^\dagger$ respectively, from which the master equation can be directly derived via Bloch-Redfield approach \cite{Jeske2013formalism, Jeskethesis, Vogt2013} which is a convenient version of the equivalent master equation techniques \cite{Breuerbook, Weissbook, Rivasbook, Gardinerbook}. 

The Bloch-Redfield master equation in $k$-space is then directly based on calculating the environmental spectral functions using the bath operators $\tilde B_{1\mathbf{k}}, \tilde B_{2\mathbf{k}}$ in the interaction picture of system and bath.  There are four environmental spectral functions due to the four combinations of the two types of bath operators. However, only two out of the four turn out to be non-zero\footnote{similar calculation see e.g.~p.50 in reference \cite{Jeskethesis}}. These are the coefficients in the master equation responsible for absorption processes and emission processes. The spectral function which corresponds to phonon absorption is given by
\begin{align}
C^{\text{abs}}_{1\mathbf{k},2\mathbf{k}'}&[-\omega(\mathbf{k})]=\int_{-\infty}^\infty d\tau e^{-i \omega(\mathbf{k}) \tau} \left \langle \tilde B_{1\mathbf{k}} (\tau) \tilde B_{2\mathbf{k}'}(0) \right \rangle\\
&=2S\int_{-\infty}^\infty d\tau e^{-i \omega(\mathbf{k}) \tau} \sum_{j,j'} g^*(\omega_j) g(\omega_{j'}) |F(k)|^2 
\left \langle \tilde A^{\dagger}_{\mathbf{k},j}(\tau) \tilde A_{\mathbf{k'},j'}(0)\right \rangle\\
&=2S\int_{-\infty}^\infty d\tau e^{-i \omega(\mathbf{k}) \tau} \sum_{j,j'} g^*(\omega_j) g(\omega_{j'}) |F(k)|^2 
\, e^{i \omega_j \tau} \bar n(\omega_j) \delta_{j,j'} \delta_{\mathbf{k}\mathbf{k}'}\\
&=2S |F(k)|^2 \sum_{j} |g(\omega_j)|^2  \delta[\omega(\mathbf{k})-\omega_j]\, \bar n(\omega_j)\\
&=2S |F(k)|^2 \mathcal{J}[\omega(\mathbf{k})] \,\bar n[\omega(\mathbf{k})] \label{eq. last step}
\end{align}
where $\bar{n}(\omega):=[\exp(\hbar \omega / k_B T)-1]^{-1}$ is the Bose-Einstein distribution and $\mathcal{J}(\omega)=\sum_j g^2(\omega_j) \delta(\omega-\omega_j)$ is the spectral density. 
In the above calculation the exponential $e^{i \omega_j \tau}$ comes from the interaction picture $\tilde A^{\dagger}_{\mathbf{k},j}(\tau)= e^{i H_{env} \tau} \,A^{\dagger}_{\mathbf{k},j}  \,e^{-i H_{env} \tau} = e^{i \omega_{\mathbf{k},j} \tau} A^{\dagger}_{\mathbf{k},j}$ due to the environmental Hamiltonian $H_{env}=\sum_{\textbf{k},j} \omega_{\mathbf{k},j} \,A^{\dagger}_{\mathbf{k},j}A_{\mathbf{k},j}$ with an environmental dispersion relation $\omega_{\mathbf{k},j}$ which as a function of $\mathbf{k}$ is given by the Fourier transform of the spatial coupling function between different environmental bosons. Consistent with phonons in the harmonic approximation, we assume uncoupled environmental bosons and hence $\omega_{\mathbf{k},j}=\omega_j$ and $\bar n(\omega_{\mathbf{k},j})=\bar n(\omega_j)$. 
In the last step of eq.~\ref{eq. last step} we simply change the argument of the Bose-Einstein distribution; this is allowed because of the delta-function.
Analogously we calculate the other spectral function, which describes emission processes:
\begin{align}
C^{\text{em}}_{2k,1k'}&[\omega(\mathbf{k})]= 2S \int_{-\infty}^\infty d\tau e^{i \omega(\mathbf{k}) \tau} \left \langle \tilde B_{2\mathbf{k}}(\tau) \tilde B_{1\mathbf{k}'} \right \rangle\\
&=2S \int_{-\infty}^\infty d\tau e^{i \omega(\mathbf{k}) \tau} \sum_{j,j'} g(\omega_j) g^*(\omega_{j'}) |F(k)|^2 
 \, e^{-i \omega_j \tau} \left \langle  A_{\mathbf{k},j} A^{\dagger}_{\mathbf{k'},j'}\right \rangle\\
&=2S \,|F(k)|^2 \mathcal{J}[\omega(\mathbf{k})] \,(\bar n(\omega_j)+1)
\end{align}

With this the master equation for the density matrix $\rho$ becomes (note that $\hbar=1 $ throughout):
\begin{multline}
\dot \rho = i[\rho, H_s] 
+ 2S \sum_\textbf{k} |F(k)|^2 \mathcal{J}[\omega(\textbf{k})]\bar n_\mathbf{k} \left(a_\mathbf{k}^\dagger\rho a_\mathbf{k} - \frac{1}{2}\{a_\mathbf{k} a_\mathbf{k}^\dagger , \rho\} \right)   \\
+ 2S \sum_\textbf{k} |F(k)|^2 \mathcal{J}[\omega(\mathbf{k})](\bar n_\mathbf{k}+1) \left(a_\mathbf{k}\rho a_\mathbf{k}^\dagger - \frac{1}{2}\{a_\mathbf{k}^\dagger a_\mathbf{k} , \rho\} \right)
\label{master equation for ferromagnet in k-space}
\end{multline}
where the Bose-Einstein distribution $ \bar n_\textbf{k}$ and the spectral density $\mathcal{J}(\omega)$ are as defined below eq.~\ref{eq. last step} and $\omega(\textbf{k})=\gamma B + 2 J S\left[-D + \sum_{\mu=1}^3 \cos(k_\mu d_\mu)\right]$ is the system's dispersion relation, derived in Eq.~\eqref{eq dispersion relation}.

For solid state environments, the spectral density of the bath is usually parameterized in the continuous limit \cite{Weissbook, Leggett1987} as:
\begin{equation}
\mathcal{J}(\omega)=\alpha\omega^s\omega_c^{s-1}{\rm e}^{-\omega/\omega_c}
\label{spectral_density}
\end{equation}
where $\alpha$ accounts for the strength of the coupling and $\omega_c$ is the cut-off frequency of the bath (typically it would be the Debye frequency of the material). Typically three cases are distinguished, $s>1$ (super-Ohmic), $s=1$ (Ohmic) and $s<1$ (sub-Ohmic).

This master equation is of Lindblad-type and we can identify the coefficient as a single magnon decay rate towards thermal equilibration for each $k$ value: 
\begin{align}
\gamma_\textbf{k} = 2S \,|F(k)|^2 \mathcal{J}[\omega(\textbf{k})] 
\label{magnon decay rate cont}
\end{align}

\begin{figure}
\centering

\includegraphics[scale=1]{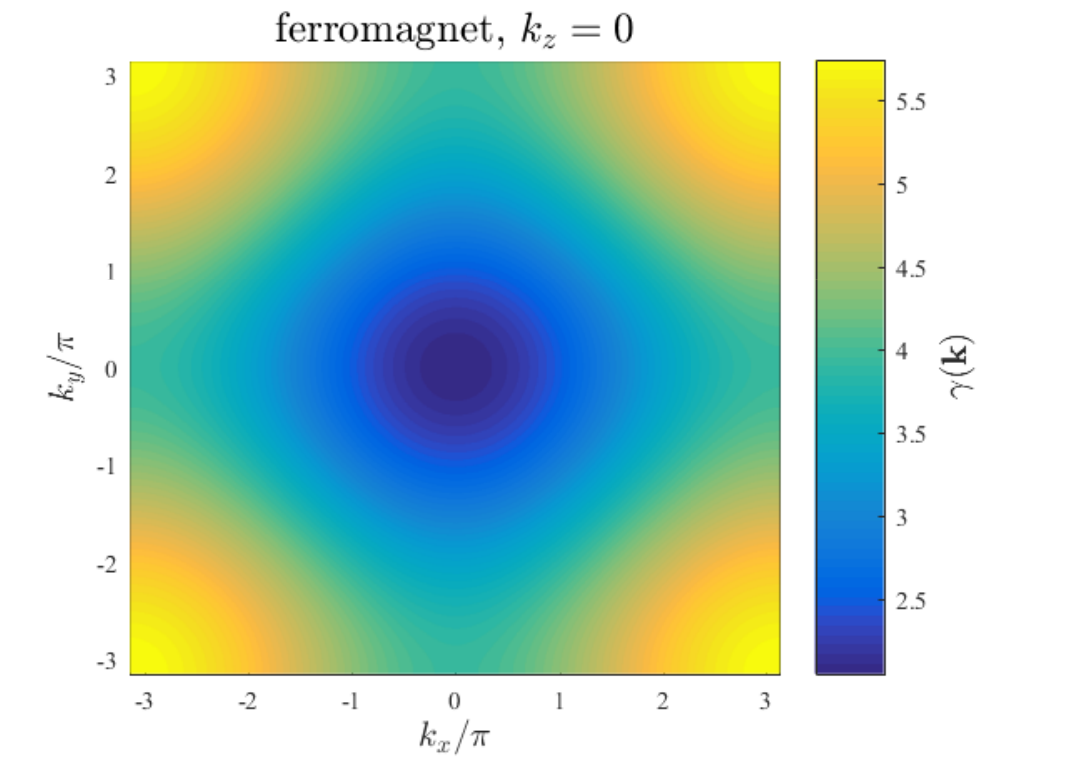}
\caption{Magnon decay rate $\gamma(\mathbf{k})$ as a function of $k_x$ and $k_y$ in the Ohmic case $s=1$, for the ferromagnetic case with $k_z=0$. The ferromagnetic case for $k_z=\pi$ looks very similar in Ohmic environments with only a slightly different $z$-axis. The noise is uncorrelated. }
\label{fig magnon decay rate uncor}
\end{figure}

We solve Eq.~\eqref{master equation for ferromagnet in k-space} for the expectation value $\langle a^\dagger_\textbf{k} a_\textbf{k} (t) \rangle$ by using the adjoint master equation in the Heisenberg picture. 
\footnote{An example of the same technique for a single harmonic oscillator can be found in section 3.4.6.2 of reference  \cite{Breuerbook}. The solution here is analogous once we keep in mind the bosonic commutation relation $[a_\textbf{k}, a_{\textbf{k}'}^\dagger]=\delta_{\textbf{kk}'}$ and details can be found in section 3.1.3 of reference \cite{Jeskethesis}.} The solution is given by
\footnote{The solution is analogous to Eq.~(3.319) in \cite{Breuerbook} and Eq.~(40) in  \cite{Rivas2013}}:
\begin{align}
\langle a^\dagger_\textbf{k} a_\textbf{k} (t) \rangle = \langle a^\dagger_\textbf{k} a_\textbf{k} (0) \rangle \; e^{-\gamma_\textbf{k} t} +  \bar n[\omega(\mathbf{k})] (1-e^{-\gamma_\textbf{k} t})
\end{align}
This result has several very important implications: We have found that magnons decay to their thermal equilibrium with a single exponential rate $\gamma_\mathbf{k}$ and we have found an analytical expression for this decay rate for any magnon with a given wave vector $\mathbf{k}$ and any given spatial noise correlation function $f(|\mathbf{u}|)$. This magnon decay rate shows how fast different plane spin waves thermalise in the Brillouin zone, and equally applies to single magnon wave packets big enough to behave like plane waves. This magnon decay rate determines their decoherence. This, together with the magnon speed, determines the coherence range of the magnons and is an important indicator for their usefulness as quantum and classical information carriers. Figure \ref{fig magnon decay rate uncor} shows the magnon decay rate for spatially uncorrelated noise. We will investigate the magnon decay rate for different temperatures, and noise correlation lengths below. 

Furthermore the thermal equilibrium distribution of magnons as a function of temperature is given by the long-time limit $\langle a^\dagger_\textbf{k} a_\textbf{k} (t=\infty) \rangle = \bar n[\omega(\mathbf{k})]$ which gives the average number of magnons for any wave vector $\mathbf{k}$. It is obtained by inserting the system's dispersion relation $\omega(\mathbf{k})$ (which is itself defined by the Fourier transform of the spatial homogeneous spin-spin coupling) into the Bose-Einstein distribution $\bar n(\omega)$.

\section{Magnetisation}
Beyond the investigation of single magnons, our master equation approach also allows us to calculate the time evolution of the average magnetisation of an entire crystal. We will sum over all sites and obtain the magnetisation of the entire ferromagnet rather than the microscopic expectation value of a single site. The ability to obtain the dynamics of a macroscopic solid-state quantity from calculations of a microscopic quantum master equation shows the versatility of this master-equation approach which can be used beyond the usual applications of nanoscale quantities. The magnetisation is furthermore an experimentally more accessible quantity than the time evolution of single magnons. 

The macroscopic magnetisation $\langle m_z \rangle$ is converted to bosonic operators via the Holstein-Primakoff transform:
\begin{align}
\langle m_z \rangle&= \frac{1}{N}\sum_{\textbf{r}} \langle S^z_\textbf{r}\rangle= S-\frac{1}{N} \sum_\textbf{r} \langle a^\dagger_\textbf{r} a_\textbf{r}\rangle\\
&= S-\frac{1}{N} \sum_\textbf{k} \langle a^\dagger_\textbf{k} a_\textbf{k}\rangle
\end{align}
Accordingly the time-evolution of a non-equilibrium state of the magnetisation is given by:
\begin{align}
\langle m_z (t) \rangle = S - \frac{1}{N} \sum_\textbf{k} \left[ e^{-\gamma_\textbf{k} t} \langle a_\textbf{k}^\dagger a_\textbf{k}(0)\rangle + \bar n[\omega(\mathbf{k})] (1-e^{-\gamma_\textbf{k} t}) \right]
\end{align}
This describes the relaxation of the system for example from one (system) temperature to another (bath) temperature or from one magnetic field to another. Assuming that we start from an equilibrium state with some magnetic field $B_0 \neq B$ and then evolve to the equilibrium state with magnetic field $B$, we have $\langle a_\textbf{k}^\dagger a_\textbf{k}(0)\rangle=\bar n[\omega_0(\mathbf{k})]=\bar n[ \gamma B_0 + 2 J S \sum_{\mu=1}^3 (\cos \,k_\mu d_\mu- 1)]$.
\begin{align}
\langle m_z (t) \rangle = S - \frac{1}{N} \sum_{\textbf{k}} \bar n[\omega(\mathbf{k})] 
- \sum_{\textbf{k}} \left( \langle a_\textbf{k}^\dagger a_\textbf{k}(0)\rangle- \bar n[\omega(\mathbf{k})] \right) e^{-\gamma_\textbf{k} t}
\end{align}
The discrete summations are explicitly given by a summation over the Brillouin zone, where each $k_\mu$ (where $\mu=x,y,z$) is discretised as $k_{\mu,n}=\frac{2\pi}{N_\mu d_\mu}n$ with the whole numbers $n \in \left\{ -\frac{N_\mu}{2}, \dots,\frac{N_\mu}{2}-1 \right\}$. For large enough $N$ we can change the summation to an integral over the Brillouin zone volume:
\begin{align}
\langle m_z (t) \rangle &= S - \frac{1}{N}\sum_{k_x=-\pi/d}^{\pi/d} \sum_{k_y=-\pi/d}^{\pi/d} \sum_{k_z=-\pi/d}^{\pi/d} 
\left\{ \bar n[\omega(\mathbf{k})] +\left( \langle a_\textbf{k}^\dagger a_\textbf{k}(0)\rangle- \bar n[\omega(\mathbf{k})] \right) e^{-\gamma_\textbf{k} t} \right\}
\label{magnetisation discrete}\\
\langle m_z (t) \rangle &\approx S - \frac{d^3}{(2\pi)^3} \int_{-\pi/d}^{\pi/d} dk_x \int_{-\pi/d}^{\pi/d} dk_y \int_{-\pi/d}^{\pi/d} dk_z  \; 
\left\{ \bar n[\omega(\mathbf{k})] +\left( \langle a_\textbf{k}^\dagger a_\textbf{k}(0)\rangle- \bar n[\omega(\mathbf{k})] \right) e^{-\gamma_\textbf{k} t} \right\}
\label{magnetisation cont}
\end{align}
As an example we choose as the initial state the equilibrium distribution which would occur at high external magnetic field $\langle a_\textbf{k}^\dagger a_\textbf{k}(0)\rangle = \bar n [\omega (\mathbf{k})]$ and regard the decay to a new equilibrium position for low magnetic field. For a saturated, i.e.~fully magnetised initial state $\langle a_\textbf{k}^\dagger a_\textbf{k}(0)\rangle = \lim_{B \rightarrow \infty} \bar n [\omega (\mathbf{k})] = 0$ because the magnetic field $B$ aligns all spins to the ground state of zero magnons present. In other words the magnetic field [by appearing in the dispersion relation, Eq.~\eqref{eq dispersion relation}] effectively increases the energy offset of the system energies so that there are no excited magnon states populated in the thermal equilibrium of infinite magnetic field. This simplifies the initial magnetisation $\langle m_z (t) \rangle=S$ and the time-dependent magnetisation: 
\begin{align}
\langle m_z (t) \rangle &\approx S - \frac{d^3}{(2\pi)^3} \int_{-\pi/d}^{\pi/d} d^3k \; 
\bar n[\omega(\mathbf{k})] \left( 1- e^{-\gamma_\textbf{k} t} \right) 
\end{align}

The continuous limit for the calculation of the magnon decay rate Eq.~\eqref{magnon decay rate cont} and the magnetisation Eq.~\eqref{magnetisation cont} facilitates finding an analytical expression. The change from discrete to continuous is justified by smoothness, and smoothness in $k$-space is guaranteed by large enough number of spins $N$ in real space as this means `high resolution' in the Brillouin-zone in $k$-space. 

However we need to keep in mind, that the Fourier-transform of the spatial correlation function, Eq.~\eqref{correlation funct Fourier transform discrete} is a separate discrete calculation. This can only be changed to a continuous Fourier transform Eq.~\eqref{correlation function - Fourier transform F(k)} for a correlation length which is not too short (otherwise the correlation function is not smooth enough to change the sum to an integral) and not too long (otherwise the correlations do not decay over the length of the entire crystal and a finite summation  cannot be changed to an infinite Fourier transform. Uncorrelated noise and nearest-neighbour correlations are therefore investigated by discrete Fourier transformation, while functional forms are only investigated as continuous Fourier transforms for correlation lengths not shorter than the spin spacing.

Next we will regard the dynamics in the Ohmic case at different temperatures and for several different cases of spatial correlation functions $f(u)$. In each case we give the corresponding Fourier transform $F(k)$, Eq.~\eqref{correlation function - Fourier transform F(k)}, the magnon decay rate $\gamma_\textbf{k}$, Eq.~\eqref{magnon decay rate cont}, and a numerical plot of the decay of the magnetisation to its equilibrium value, Eq.~\eqref{magnetisation cont}. Note that most variables are dimensionless in the following and the plots hence make relative statements about the effects of spatial correlations. For those spatial correlation functions with a correlation length $\xi$ this length is given in units of the spacing $d$ between the system's (equally spaced $d=d_x=d_y=d_z$) sites, i.e.~$\xi=10$ means a correlation length of 10 spins.

In the Ohmic case the spectral density, Eq.~\eqref{spectral_density}, becomes:
\begin{align}
\mathcal{J}[\omega(\mathbf{k})] &= \alpha \omega(\mathbf{k}) {\rm e}^{-\omega(\mathbf{k})/\omega_c} \qquad \\
& \text{with }\;\; \omega(\textbf{k})=2 J S\left[-3 + \sum_{\mu=1}^3 \cos(k_\mu d_\mu)\right]
\end{align}
For a sufficiently high cut-off frequency ($\omega_c=100 \, |J|$ in the following calculations), the spectral density as a function of $\mathbf{k}$ is essentially proportional to the system's dispersion relation. Note that a large cut-off frequency $\omega_c$ does not contradict a weak system-environment coupling $g(\omega_j)$ mentioned in section \ref{sec Interaction with a thermal environment}.

\subsection{Thermal noise in spatially uncorrelated environments}

For spatially uncorrelated thermal noise we have:
\begin{align}
f(u)&= \delta^{(3)}_{\textbf{u},0}=\gamma_0\; \delta_{u_x,0} \;\delta_{u_y,0}\; \delta_{u_z,0}\\
F(k)&=1 \\
\gamma_\textbf{k}&=  2S \,|F(k)|^2 \mathcal{J}[\omega(\textbf{k})] \\
&= 4 S^2 \alpha J \left[-3 + \sum_{\mu=1}^3 \cos(k_\mu d_\mu)\right]{\rm e}^{-\omega(\mathbf{k})/\omega_c} 
\end{align}

\begin{figure*}
\centering 
\includegraphics[scale=1]{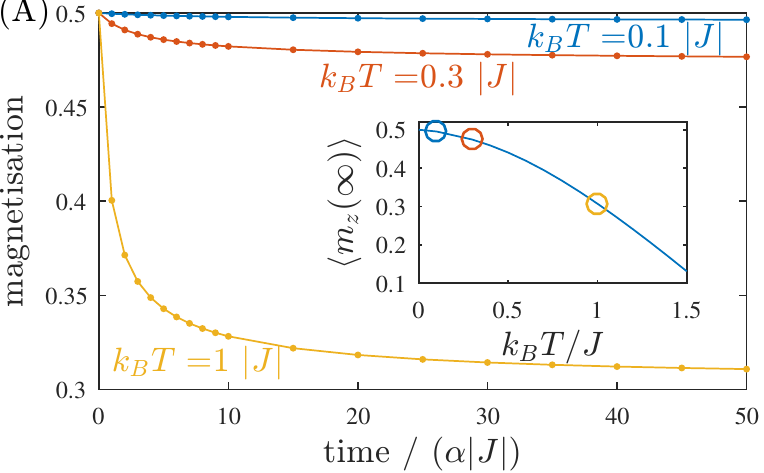}
\includegraphics[scale=1]{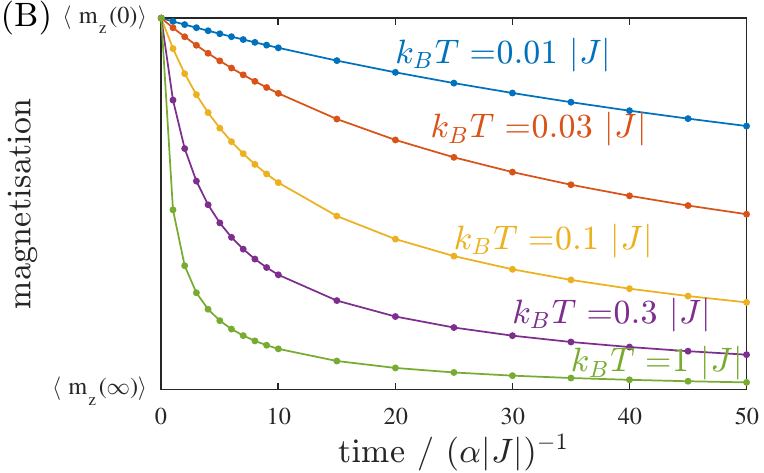}
\caption{(A): Thermal noise at different temperatures relative to the spin-spin coupling strength $J$ for spatially uncorrelated noise. The magnetisation decays from saturation to the thermal equilibrium $\langle m_z(\infty)\rangle$, which has reduced magnetisation with increasing temperature as expected. \;\; (B): The normalised decay from initial to final magnetisation shows furthermore a slightly faster decay with increasing temperature. The magnetisation is in reduced units corresponding to an ensemble average $\langle\sigma_z \rangle $, i.e. 0.5 being a fully magnetised ferromagnet.}
\label{fig magnetisa thermal noise}
\end{figure*}

Figure \ref{fig magnetisa thermal noise} shows the decay of magnetisation with time from a fully magnetised ferromagnet to its thermal equilibrium in the absence of external magnetic field. The speed of the dynamics is governed by the noise intensity, which is set by the energy exchange between system and environment $\alpha$ (i.e.~the square of their coupling) as well as $J$, since in the Ohmic case regarded here, the spectral density scales linearly with frequency $\omega(\bf{k})$, which in turn is linear in $|J|$ in the absence of external magnetic field. Thus the time is given in units of $\alpha |J|$. 

Increasing temperature decreases the equilibrium value of the magnetisation as increased thermal noise disturbs the magnetic order increasingly. For low temperatures the equilibrium value reaches close to saturation magnetisation. The relevant scale of temperature $k_B T$ is set by the system energies which in the absence of external magnetic field are defined by $\hbar |J|$. Apart from changing the equilibrium magnetisation, temperature also changes the decay slightly, which can be seen in Fig.~\ref{fig magnetisa thermal noise} by rescaling the decay curves at different temperatures to start at the same initial and final values. Increasing temperature leads to an overall slightly faster decay of the magnetisation as it increases the thermally populated phonons, which in turn increases the interactions between system and environment.

\subsection{Spatial nearest-neighbour correlations in the noise}
An uncorrelated noise environment is only found in the limit where the noise correlation length is far below the lattice constant. This is the case when the phononic noise signal varies randomly on those short length scales. We first investigate a correction to this limit, where the spatially decaying interaction between a phononic site and the spins is long enough to reach not just one spin but also the nearest neighbour spin. We therefore compare spatially uncorrelated noise with an environment where the noise of nearest-neighbours of the cubic lattice has a small correlation. This will develop a first understanding of how the introduction of any spatial correlations in the noise changes the decay characteristics of the magnetisation. We introduce the parameter $\eta$, which quantifies the amount of nearest neighbour noise correlation, where $\eta=0$ is the limit of uncorrelated noise:
\begin{align}
f(u)&=\delta^{(3)}_{\textbf{u},0}+\eta \, \delta^{(3)}_{u,1}\\
F(\textbf{k})&=1+\eta \sum_{\mu=1}^3 \cos(k_\mu d_\mu)
\end{align}

\begin{figure*}
\centering
\includegraphics[scale=1]{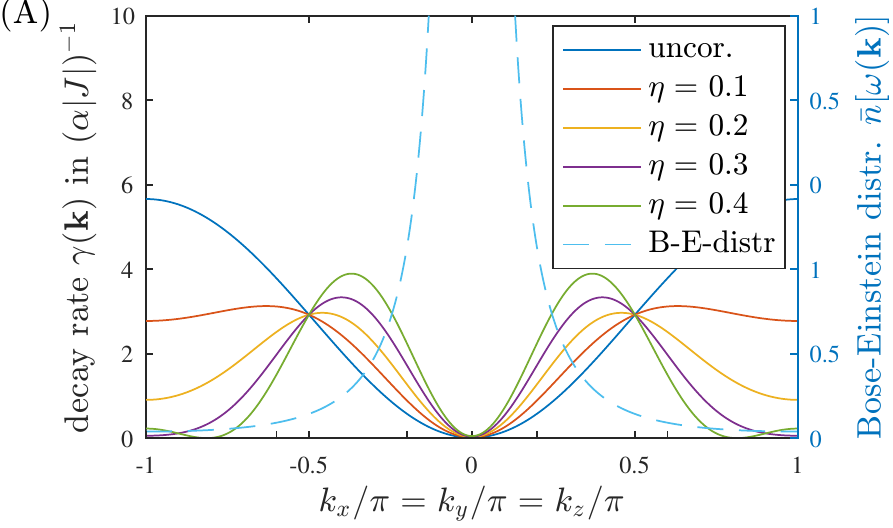}
\includegraphics[scale=1]{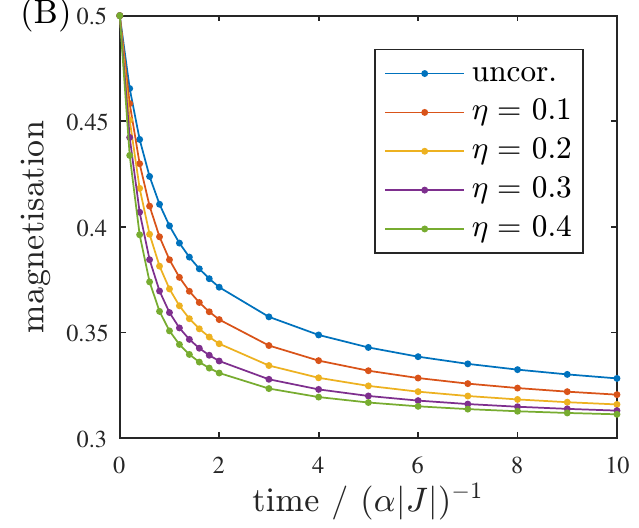}
\caption{(A): The magnon decay rate $\gamma(\bf{k})$ for uncorrelated and nearest-neighbour correlated noise on the diagonal through the cubic Brillouin zone $k_x=k_y=k_z$. With the introduction of nearest-neighbour correlations we see an increase in the decay rate for small $\mathbf{k}$-values and a reduction for high $\mathbf{k}$-values. The Bose-Einstein distribution of the magnons is heavily weighted towards lower $\mathbf{k}$-values, thus the increase in decay rate is the dominating effect in the time evolution. \;\;(B): The magnetisation decays faster for nearest-neighbour correlated noise. The temperature was set $k_B T= |J|$, the nearest-neighbour correlation strength to $\eta=0.2$.}
\label{fig magnetisa n.n. noise vs uncor}
\end{figure*}

Figure \ref{fig magnetisa n.n. noise vs uncor}A shows the change to the magnon decay rate by introducing nearest-neigbour correlations in the noise: The decay rate increases for low $k$-values, while it reduces for high $k$-values, i.e. for fast magnons. This can be useful in the context of magnonics, where typically fast magnons are used for information transport and slow decay is desirable. For the change of the magnetisation from one equilibrium state to another the increased decay rate will be the dominant effect since the Bose-Einstein distribution is heavily weighted towards low $k$-values. Indeed Fig. \ref{fig magnetisa n.n. noise vs uncor}B shows that the magnetisation decays faster to its equilibrium value with increasing nearest-neighbour correlations in the noise than in the uncorrelated case. The magnon decay rates in the Brillouin zone change considerably with the introduction of noise correlations. At lower values the decay rates increase, which is the strongest effect as the thermal distribution is weighted towards $\mathbf{k}=0$. At the edges of the Brillouin zone however spatial noise correlations strongly reduce the decay rates of fast magnons. This leads to a maximum of the decay rate for intermediate $\mathbf{k}$-values.

\subsection{Gaussian spatial decay of noise correlations}
Rather than the simple but unrealistic nearest-neighbour correlations, we now consider Gaussian spatial decay of noise correlations with a correlation length $\xi$. This also allows us to investigate much stronger spatial correlations, since correlations between all pairs of spins are considered and since a correlation length of $\xi=1.3$ spin spacings already implies that nearest neighbours have a correlation above 0.5. The spatial correlation function is:
\begin{align}
f(u)&=\exp(-|\textbf{u}|^2/\xi^2)\\
F(k)&=\pi^{3/2} \xi^3 \exp\left(-\frac{k_x^2+k_y^2+k_z^2}{4} \xi^2\right)
\end{align}
We find that $F(k)$ and hence the magnon decay rate becomes peaked around zero with a peak width inversely proportional to the correlation length $\xi$. 

\begin{figure}
\centering
\includegraphics[scale=1]{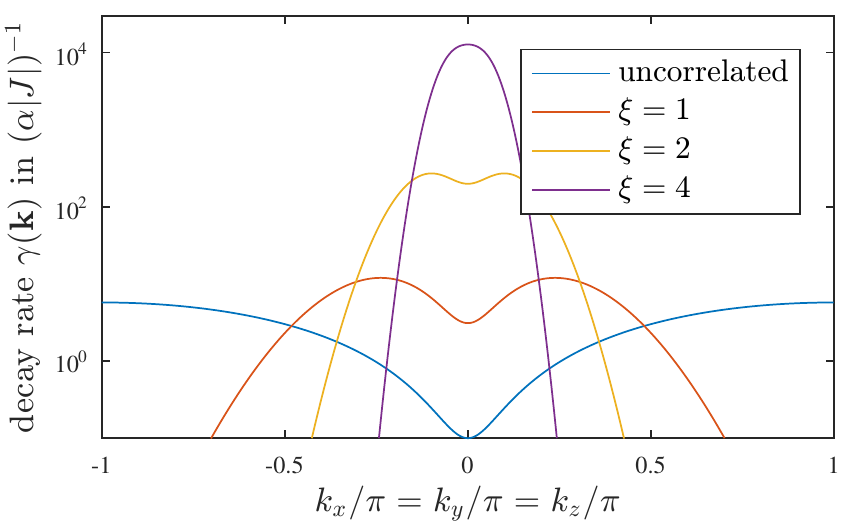}
\caption{Magnon decay rate $\gamma(\mathbf{k})$ in the Brillouin zone for uncorrelated noise and Gaussian noise correlations with different correlation length $\xi$. With more correlations in the noise the decay rate increases strongly at low $k$-values (super-radiance) while it decreases for high $k$-values (sub-radiance).}
\label{fig Gaussian Gamma}
\end{figure}

\begin{figure}
\centering
\includegraphics[scale=1]{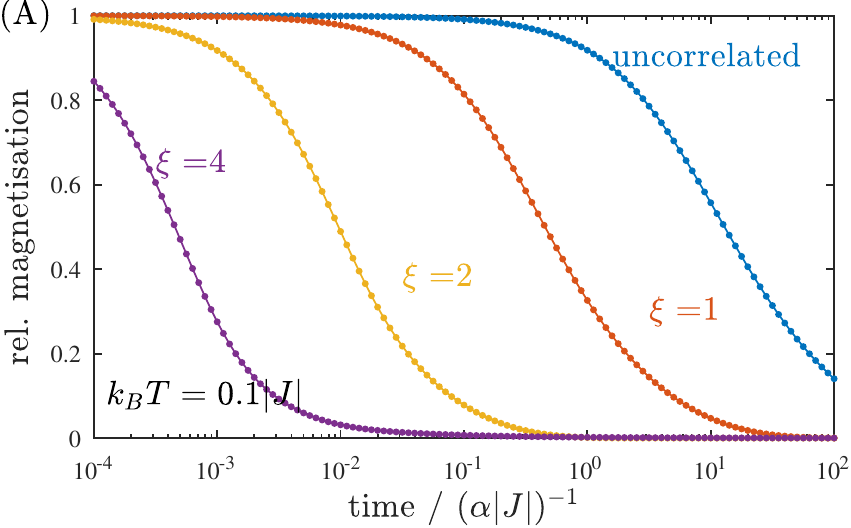}
\includegraphics[scale=1]{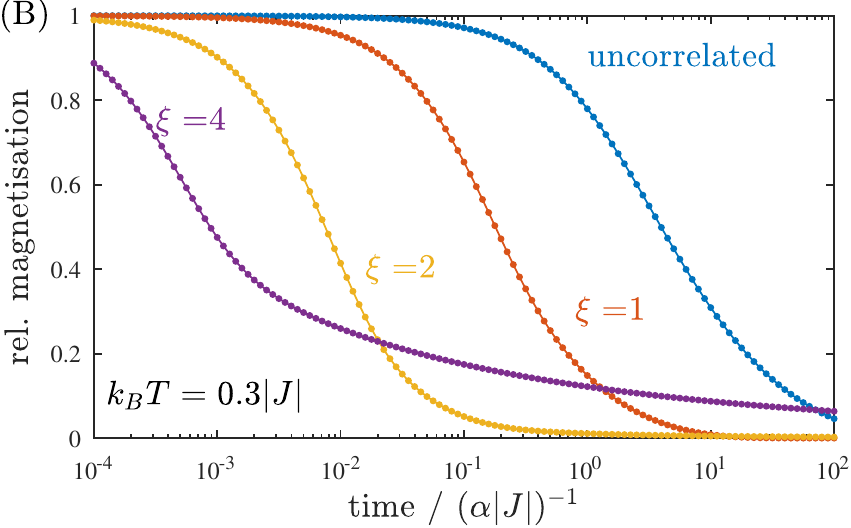}
\includegraphics[scale=1]{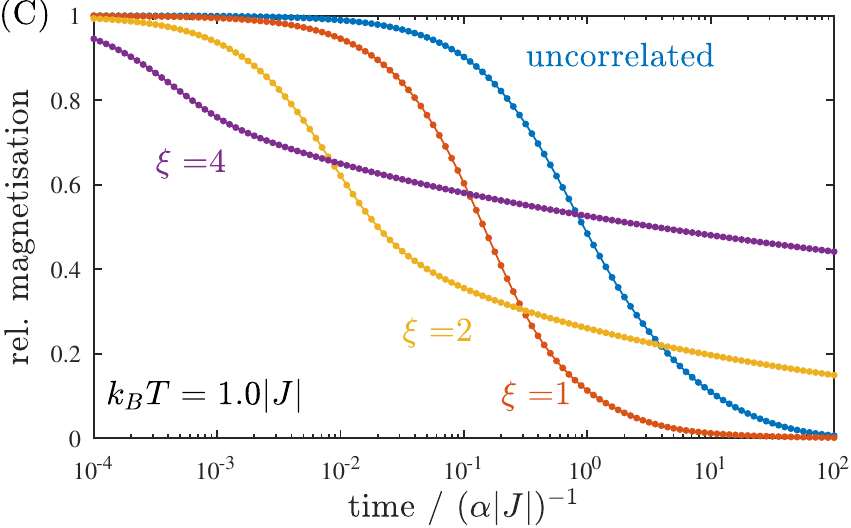}
\caption{Decay of a fully magnetised 3D-ferromagnet to the thermal equilibrium for different correlation lengths $\xi$ at (A) low temperature, (B) medium temperature and (C) high temperature. Note that all curves start at relative magnetisation 1 for $t=0$, but due to logarithmic scaling we cut off time scales below $(\alpha |J|)^{-1}=10^{-4}$.}
\label{fig magnetisa Gaussian}
\end{figure}

Figure \ref{fig Gaussian Gamma} shows the magnon decay rate for spatial noise correlations with different correlation lengths $\xi$. Since a strong increase of the decay rate is observed for small $k$-values we have plotted the $y$-axis with logarithmic scaling. We can see that the decay rate very close to $\mathbf{k}=0$ increases very strongly and the fact that this same decay rate remains close to zero in the case of only nearest-neighbour correlations (Fig.~\ref{fig magnetisa n.n. noise vs uncor}A) seems to be resulting from the nearest-neighbour restriction of noise correlations.

We have seen that the introduction of spatial correlations and the increase of the correlation length in the phononic noise of a quantum ferromagnet leads to an increased magnon decay for small wave vectors and a decreased decay for large wave vectors in the Brillouin zone. It is interesting to point out that within the approximations of our model the introduction of \emph{spatial} noise correlations, can be entirely reflected in $k$-space only by a change in the decay rate $\gamma_k$ while different $k$-modes do not experience correlated noise. This shows that fundamentally noise correlations are dependent on the operator basis and that a Fourier lattice transform can mathematically change a system from experiencing correlated noise to uncorrelated noise. This is mathematically parallel to the fact that the system's spatial couplings (i.e.~off-diagonal elements in the Hamiltonian) are only reflected in the dispersion relation of the diagonal $k$-space Hamiltonian, which does mathematically not show any `coupling' between different $k$-modes of magnons.

The effect of both increased and decreased decay rates due to spatial noise correlations in a Heisenberg-model ferromagnet can be considered an extension to the effect of super- and sub-radiance in an atomic gas, first discussed by Dicke \cite{Dicke1954}. Super- and sub-radiance are phenomena with increasingly rich relevance and applications: Sub-radiance due to correlated noise has been shown to enable undisturbed classical information transport in spin chains \cite{Jeske2013spinchain}. Recently super-radiance has been measured in ensembles of nitrogen-vacancy centres \cite{Juan2017}. 

\begin{figure*}
\centering
\includegraphics[scale=1]{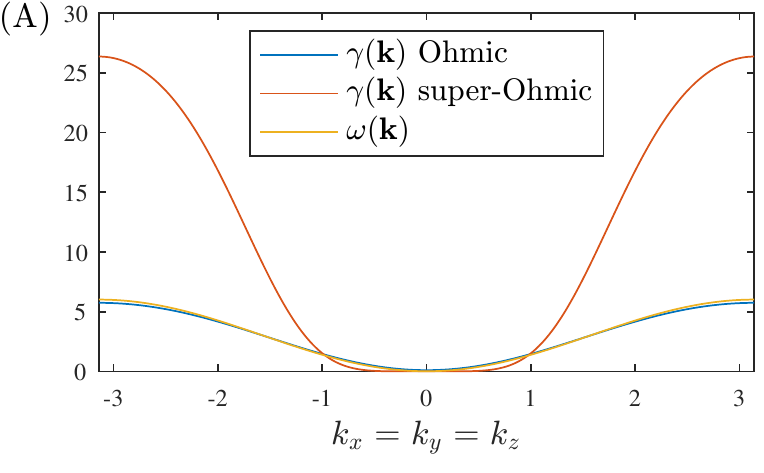}
\includegraphics[scale=1]{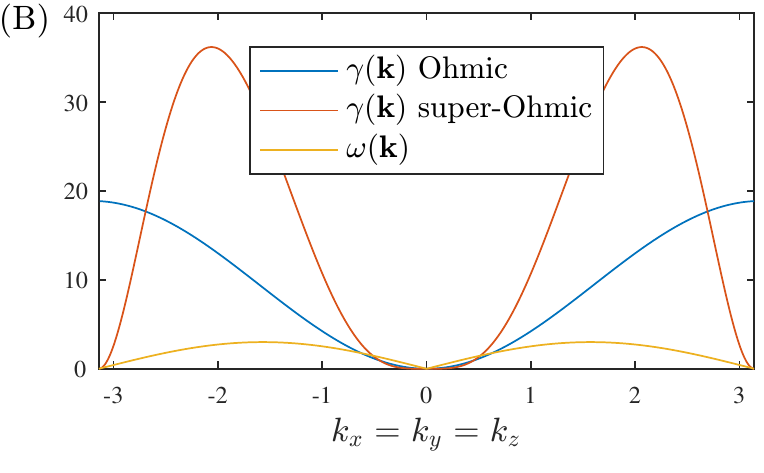}
\caption{(A): Dispersion relation $\omega(\mathbf{k})$ and magnon decay rate $\gamma(\mathbf{k})$ for a ferromagnet \;\; (B): the same for an antiferromagnet.}
\label{fig Ohmic Super-Ohmic comparison}
\end{figure*}

\begin{figure}[bth]
\centering
\includegraphics[scale=1]{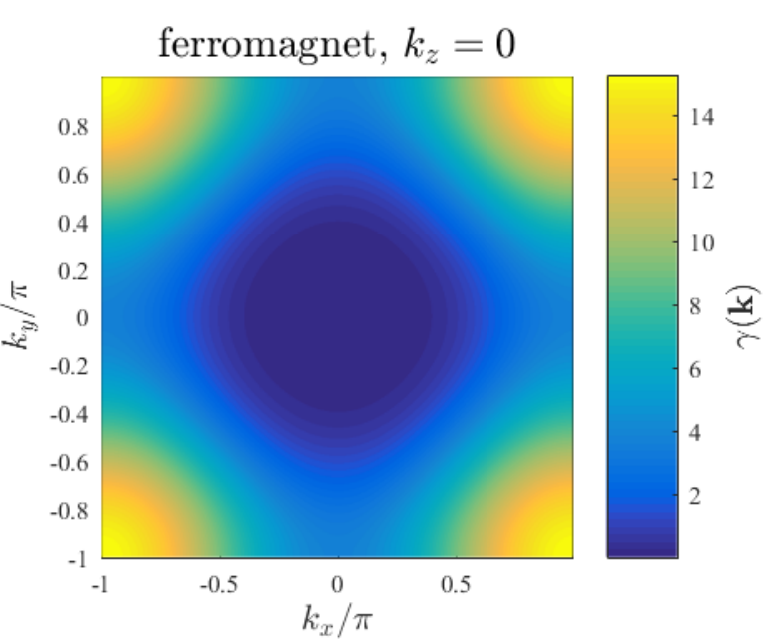}
\includegraphics[scale=1]{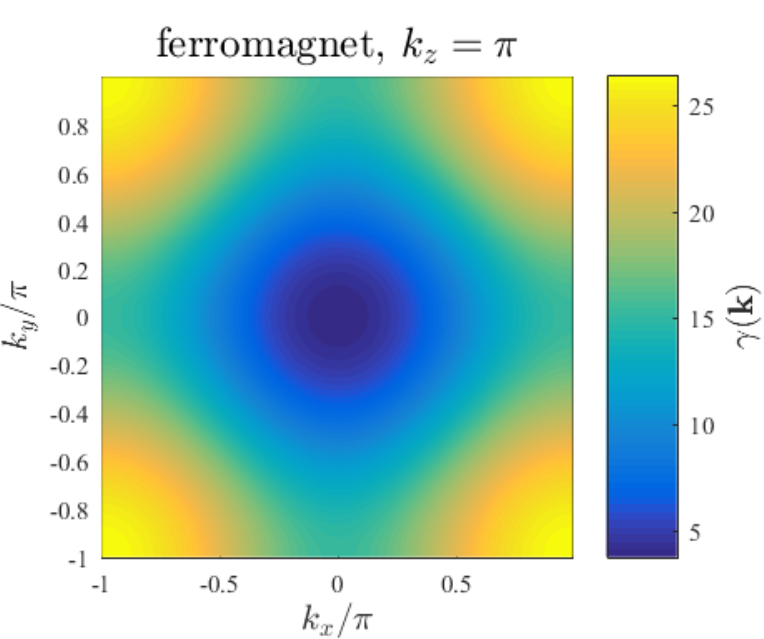}
\includegraphics[scale=1]{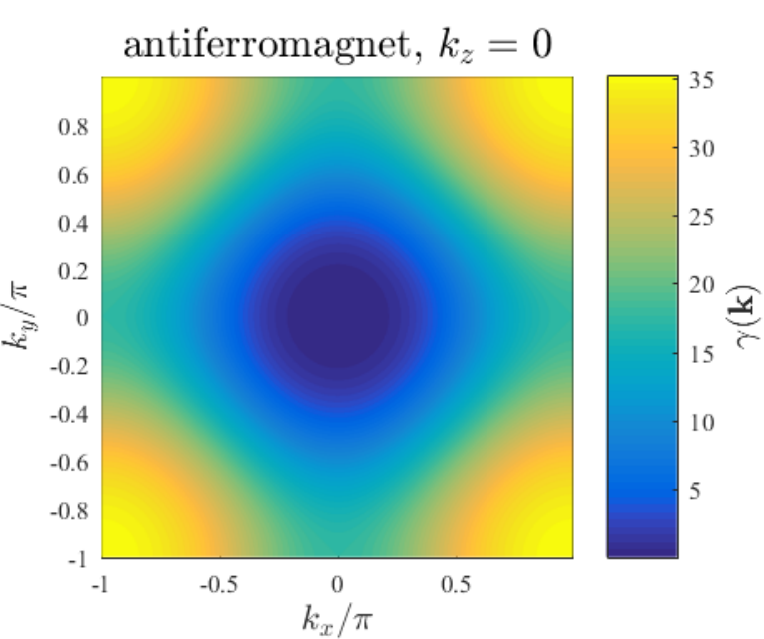}
\includegraphics[scale=1]{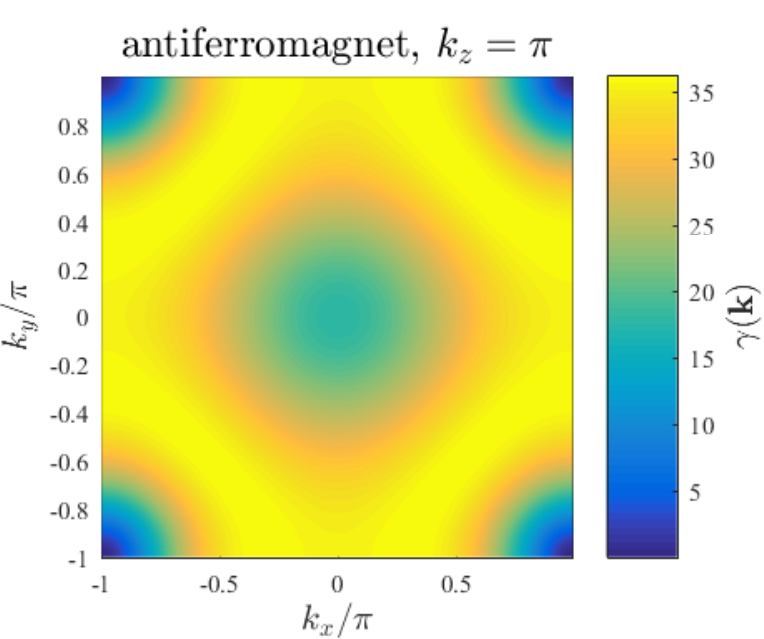}
\caption{Magnon decay rate $\gamma(\mathbf{k})$ as a function of $k_x$ and $k_y$ in the super-Ohmic case s=3, for the ferromagnetic case (top) and antiferromagnetic case (bottom) with $k_z=0$ (left) and $k_z=\pi$ (right)}
\label{fig Ferro Antiferro 3D plots of gamma(k)}
\end{figure}

The occurrence of super-radiance around small wave vectors and sub-radiance for large wave vectors can be understood by regarding the single-magnon states which are given by (see Eq.~(7.244) in reference  \cite{Noltingmagnetism}) $\ket{\mathbf{k}}=\sum_r e^{ikr}\ket{r}$, where $\ket{r}$ represents the state with only the one spin at position $r$ flipped against the external magnetic field and all others aligned. These states are the eigenvectors of the Hamiltonian and correspond to the decay rate $\gamma_k$. We can therefore regard these states at small and large $k$ values as corresponding to the occurrence of super- and sub-radiance respectively in the ferromagnet. These single magnon states in the limit of $k\approx 0$ and $k \approx \pm \pi/d$ become very similar to the prototypical super- and sub-radiant states $\ket{\uparrow \downarrow}+\ket{\downarrow \uparrow}$ and  $ \ket{ \uparrow \downarrow }-\ket{\downarrow \uparrow}$ respectively. The sub-radiance originates from the relative phase of $e^{ikr}=e^{i \pi}=-1$ which leads to a cancellation in the environmental coupling. In this context it is also interesting to mention that the introduction of classical noise into a quantum evolution has been associated with sustaining a broader momentum distribution in the time evolution \cite{Li2015}. 

Figure \ref{fig magnetisa Gaussian} shows the magnetisation as a function of time for different temperatures and correlation lengths. Since the equilibrium magnetisation $\langle m_z(t=\infty)\rangle$ is dependent on the temperature as shown in the inset of Fig.~\ref{fig magnetisa thermal noise}A, we plot the relative magnetisation $\mathcal{M}$ to focus on the spatial correlation effects and compare these at different temperatures.
\begin{align}
\mathcal{M}=\frac{\langle m_z(t)\rangle - \langle m_z(\infty)\rangle}{\langle m_z(0)\rangle - \langle m_z(\infty)\rangle}
\end{align}
At low temperatures we find a similar result to nearest-neighbour correlations: increasing correlation length leads to a faster decay of the correlations, i.e. super-radiance is dominating. At higher temperatures however, we start to notice a second effect: there is also a very slowly decaying part, which becomes increasingly important with longer correlation length. Correlated noise pushes the magnetisation decay rate into both extremes: a fast decaying part and a slowly decaying part. In Fig.~\ref{fig Gaussian Gamma} we saw that super-radiance occurs around $\mathbf{k}=0$ and subradiance occurs for large $k$-values. At low temperatures, where the distribution of magnons is peaked at zero, it therefore makes sense that this super-radiant effect is dominant. At higher temperatures, where the distribution of magnons is wider the sub-radiant effect becomes more prominant. For longer correlation lengths, where the magnon decay rate is peaked with a smaller width eventually subradiance dominates at high temperatures. 

\section{Comparing ferromagnet and antiferromagnet}
The antiferromagnet has been discussed in detail in ref.~\cite{Rivas2013}. It can be calculated in a similar way, however an additional Bogoliubov transformation is required after the Holstein-Primakoff transform. The resulting dispersion relations and magnon decay rates are different from the ferromagnetic case. We will show below a short comparison in uncorrelated noise environments and want to point out that this comparison is strongly dependent on the type of spectral density. For Ohmic spectral density the magnon decay rate in the Brillouin zone behaves qualitatively the same. Figure \ref{fig magnon decay rate uncor} showed the ferromagnetic case and Fig.~\ref{fig Ohmic Super-Ohmic comparison} compares this to the (in the Ohmic case) similar antiferromagnetic behaviour. 

\subsection{Ohmic vs Super-Ohmic spectral densities}
Figure \ref{fig Ohmic Super-Ohmic comparison} compares the decay rates of the super-Ohmic and Ohmic case and allows a comparison to the dispersion relation. For ease of display we have plotted all functions as a function of a diagonal line through the Brillouin zone. While the magnon decay rate as a function of the wave vector $\mathbf{k}$ behaves quite similar in Ohmic spectral density, this is no longer the case for super-Ohmic spectral densities, where the magnon decay rate in the antiferromagnetic case goes to zero for magnons with maximal $\mathbf{k}$. Figure \ref{fig Ferro Antiferro 3D plots of gamma(k)} shows this behaviour with two $k_z$-slices of the Brillouin zone. This strongly reduced decay rate of fast magnons in the antiferromagnet is an interesting result. In the context of quantum information transport a fast magnon with a slow decay rate means that information might be transported robustly. Note that this is true independent of temperature and the number of spins.

The cause of the different behaviour lies in the antiferromagnetic dispersion relation, which goes to zero for magnons at maximal $\mathbf{k} =(\pm \pi/d_x,\pm \pi/d_y,\pm \pi/d_z)$. The antiferromagnetic dispersion relation is given by \cite{Rivas2013}:
\begin{align}
\omega_{\text{AF}}(\mathbf{k})=2 J_{\text{AF}} S \sqrt{D^2 - \left( \sum_{\mu=1}^3 \cos k_\mu d_\mu \right)^2}
\end{align}
This behaviour translates to the decay rate due to the spectral density in the super-Ohmic case. In the Ohmic case the behaviour of the dispersion relation does not translate to the decay rates because the linear scaling of the dispersion relation is counteracted by an additional factor in the antiferromagnetic magnon decay rate \cite{Rivas2013} which scales as $(D+\sum_\mu k_\mu d_\mu )^{-1}$ and leads to finite values of the decay rate for maximal $\mathbf{k}$ in the Ohmic case.

\section{Conclusion}
In conclusion, we found an analytical solution for the time-evolution of magnons in a quantum ferromagnet with spatially correlated phononic noise. The magnons decay with a single exponential decay rate to their thermal equilibrium. This remains true for both spatially uncorrelated and correlated phononic noise environments and we have found an analytical expression for this decay rate as a function of wave vector $\mathbf{k}$ and the noise correlation function $f(|\mathbf{r}-\mathbf{r}'|)$. The introduction of noise correlations causes faster decay (super-radiance) of slow magnons and slow decay (sub-radiance) of fast magnons. 

We also calculated the total macroscopic magnetisation of the entire quantum ferromagnet from the microscopic master equation. The magnetisation decays from a fully magnetised state to its thermal equilibrium faster with higher temperature and shows less remaining magnetisation in the equilibrium state with higher temperature. Introducing nearest-neighbour correlations in the noise leads to an overall slightly faster decay of the magnetisation. However, considering the more general model of spatial correlations with a correlation length $\xi$ and a Gaussian profile we find that spatial correlations not only introduce faster decay of the magnetisation but for stronger correlations and higher temperatures there also appears a slowly-decaying part to the magnetisation. Thus strong spatial noise correlations split the decay of the magnetisation into both extremes: instead of one typical time scale there appears a fast and slow decaying part of the magnetisation.

In the comparison between a ferromagnet and an antiferromagnet we find qualitatively similar behaviour in the magnon decay rate in Ohmic environments. However in super-Ohmic noise environments, i.e. when the noise spectral density scale superlinear with the frequency, the antiferromagnet (in contrast to the ferromagnet) shows a magnon decay rate that goes to zero at the largest $|\mathbf{k}|$ values in the Brillouin zone, which may be an interesting regime for magnonics and quantum information transport.

\ack
We would like to thank Andrew Greentree for helpful discussions and support. This work was supported in part by the Australian Government through the Australian Research Council (ARC) under projects DP140100375, FT110100225 and CE170100026. It was also supported by computational resources provided by the Australian Government through the National Computational Infrastructure National Facility.
JJ acknowledges support by the German Bundesministerium f\"ur Bildung und Forschung BMBF. AR and MAMD acknowledge the Spanish MINECO grant FIS2015-67411, the CAM research consortium QUITEMAD+ S2013/ICE-2801, and U.S. Army Research Office through grant W911NF-14-1-0103 for partial financial support.

\appendix

\section{Master equation with the full spin-boson interaction}
\label{app:full spin-boson coupling}
Here we show in detail why we simplified the form of the system-environment interaction in the main article as $S^x X$. We take the full interaction Hamiltonian and show that the resulting master equation is equivalent.

The full interaction Hamiltonian is given by $H_{int}\propto \bm{S}\cdot \bm{R}$, where $\bm{S}=(S^x,S^y,S^z)$ is the spin vector and $\bm{R}=(X,Y,Z)$  the position operators of the bosonic environment. Here we assume a local environmental model,
\begin{eqnarray}\label{V}
H_{int}&=&\sum_j\sum_{\bm{r}} g(\omega_j)\left[S^x_{\bm{r}}(A_{\bm{r},j}^x+A^{x\dagger}_{\bm{r},j})\right.\nonumber\\
&+&\left.S^y_{\bm{r}}(A_{\bm{r},j}^y+A^{y\dagger}_{\bm{r},j})+S^z_{\bm{r}}(A_{\bm{r},j}^z+A^{z\dagger}_{\bm{r},j})\right],
\end{eqnarray}
here $A$ and $A^\dagger$ stand for annihilation and creation operators of the environmental boson modes, and we have assumed that the coupling function $g(\omega_j)$ is isotropic and the same for every member of the lattice. On the other hand, the Hamiltonian of the environment is
\begin{equation}\label{HE}
H_{env}=\sum_j \sum_{\bm{r}} \omega_j(A^{x\dagger}_{\bm{r},j}A_{\bm{r},j}^x+A^{y\dagger}_{\bm{r},j}A_{\bm{r},j}^y+A^{z\dagger}_{\bm{r},j}A_{\bm{r},j}^z),
\end{equation}
which is written as
\begin{equation}
H_{env}=\sum_j \sum_{\bm{k}} \omega_j(A^{x\dagger}_{\bm{k},j}A_{\bm{k},j}^x+A^{y\dagger}_{\bm{k},j}A_{\bm{k},j}^y+A^{z\dagger}_{\bm{k},j}A_{\bm{k},j}^z),
\end{equation}
after taking Fourier transform.

After Holstein-Primakoff transformation with linear spin wave approximation, the interaction term reads
\begin{align}
H_{int}&=\sum_j g(\omega_j)\sum_{\bm{r}\in A} \left[\sqrt{2S}(a_{\bm{r}}+a^\dagger_{\bm{r}})(A_{\bm{r},j}^x+A^{x\dagger}_{\bm{r},j})\right.\nonumber\\
&-{\rm i}\sqrt{2S}(a_{\bm{r}}-a^\dagger_{\bm{r}})(A_{\bm{r},j}^y+A^{y\dagger}_{\bm{r},j})\\
&+\left. (S-a^\dagger_{\bm{r}}a_{\bm{r}})(A_{\bm{r},j}^z+A^{z\dagger}_{\bm{r},j})\right]. \nonumber
\end{align}
We now can consider several simplifications of this Hamiltonian based on the following facts,
\begin{itemize}
\item The term $a^\dagger_{\bm{r}}a_{\bm{r}}$ is negligible in comparison to the others in the regime where the spin wave theory is valid: $\langle a^\dagger_{\bm{r}}a_{\bm{r}}\rangle\ll 2S$.
\item We ignore the terms $S(A_{\bm{r},j}+A^{\dagger}_{\bm{r},j})$ because they are fast oscillators in a rotating-wave approximation which we may neglect in the weak coupling limit.
\end{itemize}
Therefore, we arrive at
\begin{equation}\label{VLSW}
H_{int}=\sqrt{2S}\sum_j\sum_{\bm{r}} g(\omega_j)\left[a_{\bm{r}}(A^x_{\bm{r},j}+A^{x\dagger}_{\bm{r},j}-{\rm i}A^y_{\bm{r},j}-{\rm i}A^{y\dagger}_{\bm{r},j})+ a^\dagger_{\bm{r}}(A^x_{\bm{r},j}+A^{x\dagger}_{\bm{r},j}+{\rm i}A^y_{\bm{r},j}+{\rm i}A^{y\dagger}_{\bm{r},j})\right],
\end{equation}
and by taking the Fourier transform this interaction becomes
\begin{equation}\label{VLSWkfinal}
H_{int}=\sqrt{2S}\sum_j\sum_{\bm{k}} g(\omega_j)\left[a_{\bm{k}}(A^x_{-\bm{k},j}+A^{x\dagger}_{\bm{k},j}-{\rm i}A^y_{\bm{-k},j}-{\rm i}A^{y\dagger}_{\bm{k},j})+ a^\dagger_{\bm{k}}(A^x_{\bm{k},j}+A^{x\dagger}_{\bm{-k},j}+{\rm i}A^y_{\bm{k},j}+{\rm i}A^{x\dagger}_{\bm{-k},j})\right],
\end{equation}
Because of the Riemann-Lebesgue lemma \cite{libro}, for small coupling $g(\omega_j)$ we can safely neglect the counter-rotating terms in (\ref{VLSWkfinal}) and arrive at:
\begin{equation}
H_{int}=\sqrt{2S}\sum_j\sum_{\bm{k}} g(\omega_j)\left[a_{\bm{k}}(A^{x\dagger}_{\bm{k},j}-{\rm i}A^{y\dagger}_{\bm{k},j}) +a^\dagger_{\bm{k}}(A^{x}_{\bm{k},j}+{\rm i}A^{y}_{\bm{k},j})\right].
\end{equation}
Now the problem becomes equivalent to a collection of uncoupled harmonic oscillators given by their operators $a_{\bm k}$, which are coupled to two sets of independent environments ($x$ and $y$) characterized by $\bm{k}$. 

We assume that the environments are in the Gibbs state at some temperature $T$,
\begin{eqnarray}
\rho_E&=&Z^{-1}{\rm e}^{-\beta H_E}\\
&=&Z^{-1}{\rm e}^{-\beta\sum_j \sum_{\bm{k}} \omega_j(A^{x\dagger}_{\bm{k},j}A_{\bm{k},j}^x+A^{y\dagger}_{\bm{k},j}A_{\bm{k},j}^y+A^{z\dagger}_{\bm{k},j}A_{\bm{k},j}^z)}\nonumber,
\end{eqnarray}
where $Z=\mathrm{Tr}\left({\rm e}^{-\beta H_E}\right)$ is the partition function with $\beta=1/k_{\rm B}T$. From now on we shall use natural units $\hbar=k_\mathrm{B}=1$.

The standard tools to obtain a master equation for a weak interaction with the environment can be found in references \cite{libro,Weiss,BrPe,GardinerZoller}. The bath correlation functions split in two different types, depending on which kind of process are associated to them, emission or absorbtion. We have
\begin{align}
C_{\rm abs}(\omega(\bm{q}),\bm{k},\bm{q})&=2S\int_{-\infty}^\infty du \sum_{i,j} {\rm e}^{-{\rm i}[\omega(\bm{q})-\omega_i] u} g(\omega_i) g(\omega_j){\rm Tr} \left[\rho_E (A^{x\dagger}_{\bm{k},i}-{\rm i}A^{y\dagger}_{\bm{k},i})(A^{x}_{\bm{q},j}+{\rm i}A^{y}_{\bm{q},j})\right]\nonumber \\
&=2S\int_{-\infty}^\infty du \sum_{j}{\rm e}^{-{\rm i}[\omega(\bm{q})-\omega_j]u} g^2(\omega_j) {\rm Tr} \left[\rho_E (A^{x\dagger}_{\bm{k},j}A^{x}_{\bm{q},j}+A^{y\dagger}_{\bm{k},j}A^{y}_{\bm{q},j})\right]\\
&=4S\int_{-\infty}^\infty du \sum_{j} {\rm e}^{-{\rm i}[\omega(\bm{k})-\omega_j]u} g^2(\omega_j) \bar{n}_{\bm{k}}(\omega_j)\delta_{\bm{k},\bm{q}},
\end{align}
where
\begin{equation}
\bar{n}_{\bm{k}}(\omega_j):=[\exp(\omega_j/T)-1]^{-1}
\end{equation}
is the number of quanta in the bath labeled by ``$\bm{k}$'' at frequency $\omega_j$.
In the continuous limit, we introduce the spectral density function, $\mathcal{J}(\omega):=\sum_jg^2(\omega)\delta(\omega-\omega_j)$ (formally), and change the sum by an integral
\begin{align}\label{Cabs}
C_{\rm abs}(\omega(\bm{k}),\bm{k},\bm{q})&=S\int d\omega \int_{-\infty}^\infty du {\rm e}^{-{\rm i}[\omega(\bm{k})-\omega]u} \mathcal{J}(\omega)\bar{n}_{\bm{k}}(\omega)\delta_{\bm{k},\bm{q}}\nonumber\\
&=2\pi S\int d\omega \delta[\omega(\bm{k})-\omega]\mathcal{J}(\omega)\bar{n}_{\bm{k}}(\omega)\delta_{\bm{k},\bm{q}}\nonumber\\
&=2\pi S\mathcal{J}[\omega(\bm{k})]\bar{n}_{\bm{k}}\delta_{\bm{k},\bm{q}},
\end{align}
where $\bar{n}_{\bm{k}}:=[\exp(\omega({\bm{k}})/T)-1]^{-1}$.

Similarly we obtain that
\begin{align}\label{Cem}
C_{\rm em}(\omega,\bm{k},\bm{q})&=\frac{S}{2}\int_{-\infty}^\infty du \sum_{i,j} {\rm e}^{{\rm i}[\omega(\bm{q})-\omega_i] u} g(\omega_i) g(\omega_j){\rm Tr} \left[\rho_E (A^{x}_{\bm{k},i}+{\rm i}A^{y}_{\bm{k},i})(A^{x\dagger}_{\bm{q},j}-{\rm i}A^{y\dagger}_{\bm{q},j})\right]\nonumber \\
&=2\pi S\mathcal{J}[\omega(\bm{k})](\bar{n}_{\bm{k}}+1)\delta_{\bm{k},\bm{q}}.
\end{align}

We can see that the inclusion of the $S^y Y$ term in the spin-boson coupling of system and environment has indeed simply led to an additional noise term analogous to the one from the $S^x X$ term. The  $S^z Z$ term has led to negligible terms in the linear spin wave regime.

\section{Further details from eq.~\eqref{eq for appendix} to \eqref{eq for appendix 2}}
\label{app: identifying the delta function}
After changing the summation index from $\textbf{r}'$ to $\textbf{u}$ the interaction Hamiltonian reads:
\begin{align}
H_{int}=\sqrt{2S} \sum_j g(\omega_j) \sum_{\textbf{k},\bf{k'}} \sum_{\bf{u}} e^{i\textbf{k}'\cdot\mathbf{u}} f(|\textbf{u}|)\,a_{\bf{k}}^\dagger A_{\textbf{k}',j} 
 \sum_\textbf{r} \frac{1}{N}e^{i(\mathbf{k'-k})\cdot\textbf{r}}  + \text{h.c.} 
\end{align}
We then identify the last summation over $\textbf{r}$ as a Kronecker-delta in each dimension by inserting $k_\mu=\frac{2\pi n_\mu}{N_\mu d_\mu}$ and $r_\mu=d_\mu \tilde n_\mu$ for each component:
\begin{align}
\sum_{\tilde n_\mu=1}^{N_\mu} \frac{1}{N_\mu} \exp\left[ i\frac{2\pi}{N_\mu}(n'_\mu-n_\mu)\tilde n_\mu \right]=\delta_{n'_\mu,n_\mu}\\
\sum_\textbf{r} \frac{1}{N}e^{i(\mathbf{k'-k})\cdot\textbf{r}} = \delta_{k'_x,k_x} \delta_{k'_y, k_y} \delta_{k'_z, k_z}
\end{align}
The Hamiltonian then becomes eq. 15 from the main article after performing the summation over $\textbf{k}'$.

\section{Further simplification of eq.~\eqref{correlation function - Fourier transform F(k)}}
\label{app simplifying Fourer transform F(k) to 1D integral}
Since the function $f(|u|)$ only depends on the magnitude $|\textbf{u}|$, we can simplify this three-dimensional Fourier transform to a one-dimensional integral using polar coordinates. The result is also sometimes called a 3D Hankel transform or 3D Fourier-Bessel transform. For this integration we choose without loss of generality the polar direction to coincide with the $\textbf{k}$ direction, such that $\textbf{k}\cdot \textbf{u}=k u \cos(\theta)$ and then simplify by substitution $v=\cos \theta$: 
\begin{align}
F(\textbf{k})&=\frac{1}{V_d}\int_0^{2\pi} \int_0^\pi \int_0^\infty f(u) e^{i k u \cos \theta} u^2 \sin \theta\, du \,d\theta\, d\phi\\
&=\frac{4\pi}{V_d k} \int_0^\infty f(u) u \sin(k u) \,du
\end{align}

\end{document}